\documentclass{article}
\usepackage{amssymb}
\usepackage[utf8]{inputenc}
\usepackage{authblk}
\usepackage{setspace}
\usepackage[margin=1.25in]{geometry}
\usepackage{graphicx}
\graphicspath{ {./figures/} }
\usepackage{subcaption}
\usepackage{amsmath}
\usepackage{lineno}
\usepackage{comment}
\usepackage{float}
\usepackage{textcomp}
\usepackage[usenames,dvipsnames]{color}
\usepackage[square,numbers]{natbib}
\usepackage{hyperref}
\DeclareUnicodeCharacter{2212}{-}
\usepackage{titlesec}
\titlelabel{\thetitle.\quad}

\title{\bf Unveiling Fine Structure and Energy-driven Transition of Photoelectron Kikuchi Diffraction}

\author[1,2]{Trung-Phuc Vo}
\author[3,4]{Olena Tkach}
\author[1]{Aki Pulkkinen}
\author[5]{Didier S\'{e}billeau}
\author[6]{Aimo Winkelmann}
\author[3]{Olena Fedchenko}
\author[3,7]{Yaryna Lytvynenko}
\author[3]{Dmitry Vasilyev}
\author[3]{Hans-Joachim Elmers}
\author[3]{Gerd Sch\"onhense}
\author[1*]{J\'{a}n Min\'{a}r}

\affil[1]{New Technologies-Research Centre, University of West Bohemia, 30100 Pilsen, Czech Republic.}
\affil[2]{Institute of Physics, Czech Academy of Sciences, Cukrovarnická 10, 162 00 Praha 6, Czech Republic.}
\affil[3]{Johannes Gutenberg-Universit\"at, Institut für Physik, 55128 Mainz, Germany.}
\affil[4]{Sumy State University, Rymskogo-Korsakova 2, 40007 Sumy, Ukraine.}
\affil[5]{Univ Rennes, CNRS, IPR (Institut de Physique de Rennes) - UMR 6251, F-35000, Rennes, France.}
\affil[6]{AGH University of Krakow, Academic Centre for Materials and Nanotechnology, Krak\'{o}w, Poland.}
\affil[7]{Institute of Magnetism of the NAS of Ukraine and MES of Ukraine, 03142 Kyiv, Ukraine.}
\affil[*]{Address correspondence to: jminar@ntc.zcu.cz}
\date{\today}

\begin{document}

\maketitle
\begin{abstract}
The intricate fine structure of Kikuchi diffraction plays a vital role in probing phase transformations and strain distributions in functional materials, particularly in electron microscopy. Beyond these applications, it also proves essential in photoemission spectroscopy (PES) at high photon energies, aiding in the disentanglement of complex angle-resolved PES data and enabling emitter-site-specific studies. However, the detection and analysis of these rich faint structures in photoelectron diffraction (PED), especially in the hard X-ray regime, remain highly challenging, with only a limited number of simulations successfully reproducing these patterns. The strong energy dependence of Kikuchi patterns further complicates their interpretation, necessitating advanced theoretical approaches. To enhance structural analysis, we present a comprehensive theoretical study of fine diffraction patterns and their evolution with energy by simulating core-level emissions from Ge(100) and Si(100). Using multiple-scattering theory and the fully relativistic one-step photoemission model, we simulate faint pattern networks for various core levels across different kinetic energies (106 eV - 4174 eV), avoiding cluster size convergence issues inherent in cluster-based methods. Broadening in patterns is discussed via the inelastic scattering treatment. For the first time, circular dichroism has been observed and successfully reproduced in the angular distribution of Si (100) 1s, revealing detailed features and asymmetries up to 31\%. Notably, we successfully replicate experimental bulk and more "surface-sensitivity" diffraction features, further validating the robustness of our simulations. The results show remarkable agreement with the experimental data obtained using circularly polarized radiations, demonstrating the potential of this methodology for advancing high-energy PES investigations.
\end{abstract}

\section{INTRODUCTION}
The pursuit of novel materials with customized properties continues to drive advances in materials science, motivated by challenges in areas such as information technology and carbon-free energy production. In many instances, structural insights are essential for understanding phenomena related to electronic and magnetic properties. The rich fine structure \cite{cios2024ambiguity,winkelmann2006perpendicular,vos2017element} in the diffraction patterns plays a crucial role in investigating phase transformations, nanoscale defects, and strain distributions in functional materials. Transmission Kikuchi diffraction, an advanced electron microscopy technique, enables high-resolution crystallographic analysis in thin films and nanostructures \cite{trimby2014characterizing,zielinski2015transmission}. It has been successfully applied to study nano-oxide and ultrafine metallic grains, providing critical insights into the grain boundary character \cite{abbasi2015application}. Building on these applications, the fine details in the patterns are also instrumental in ultrafast Kikuchi diffraction. This technique allows nanoscale probing of transverse elastic waves, revealing unit cell tilting, polarization, and resonance oscillations in materials \cite{yurtsever2011kikuchi}. Its time-resolved version is used to map the (non)linear elastic response of nanoscale graphite after ultrafast strain excitation, revealing longitudinal wave echoes and a distinct nonlinear mode attributed to localized breather motion \cite{liang2014observing}. 

Kikuchi diffraction \cite{kikuchi1928diffraction,williams2009kikuchi} is extensively studied in scanning and transmission electron microscopy (SEM and TEM) using energy sources up to 100 keV, where it appears as electron backscatter diffraction patterns \cite{schwartz2009electron}. However, detecting these rich yet faint structures in photoemission spectroscopy (PES), especially in the hard X-ray regime, remains challenging, and only a limited number of simulations have successfully reproduced the experimental results \cite{vo2024layered,fedchenko2019high,williams2012observation}. Previously, key fine features of W 3d\textsubscript{5/2} at $h\nu$= 6 keV agreed reasonably well with calculations \cite{tkach2023circular}. Another study emphasized over 50\% contrast in the spin-orbit doublets and the distinct diffractogram of sub-levels for W(110) \cite{vo2024analyzing}. In PES, the rich fine structures of diffraction patterns occur in a photon-in/electron-out technique called photoelectron diffraction (PED) \cite{medjanik2021site,winkelmann2004simulation,fedchenko2020emitter}, a specific effect related to the general case of electron emission from internal sources within a crystal \cite{fedchenko2019high}. 
In PED, the angular distribution arises from the interference of photoelectron waves in the final states. The intensity variations result from the constructive or destructive interference of coherently emitted electrons 
\cite{fadley1994photoelectron,woodruff2010surface,westphal2003study,greif2013photoelectron}. Depending on the photon energy range used, PED can be classified as ultraviolet-PED (UPD) or X-ray-PED (XPD), corresponding to different photoelectron kinetic energies. PED studies can focus on either core-level electrons (CL-PED) or valence-band electrons (VB-PED). 
At lower kinetic energies, typically below ~300 eV, strong backscattering occurs \cite{fadley1994photoelectron,fadley1987photoelectron}, which can be used to extract structural information about atoms located "behind" the emitter from the detector’s perspective \cite{kaduwela1991application}. In contrast, when the kinetic energy reaches approximately 500 eV or higher, forward scattering (zeroth-order scattering) becomes the dominant mechanism \cite{thompson1984x,fadley1994photoelectron}. In this regime, the diffraction pattern is primarily distinguished by features resembling "Kikuchi patterns," including characteristic lines and bands commonly observed in high-energy electron diffraction. At high photon energies, rich fine structures of the core levels \cite{babenkov2019high,schonhense2020momentum} serves a pivotal tool to disentangle XPD effects \cite{gray2011probing,gray2012bulk,kalha2021hard,arab2019electronic,gray2013momentum}, which significantly complicate the data interpretation of angle-resolved photoemission spectroscopy (ARPES).

The interplay between kinetic energy and inelastic scattering governs the resolution and contrast of the diffraction patterns seen in photoelectron diffraction experiments \cite{fadley2010x,fadley2014some,winkelmann2014influence}. At low kinetic energies ($<$ 50 eV), electrons undergo strong inelastic scattering due to their short inelastic mean free path (IMFP). 
This results in significant intensity attenuation and diffuse Kikuchi bands, causing the diffraction features to appear broadened and less structured. Another reason is that scattering amplitude shows strong angular dependency. As the kinetic energy increases to the intermediate range (100–500 eV), the IMFP extends to a few nanometers, 
inelastic scattering effects are reduced and diffraction peaks become more distinguishable. However, some broadening persists, particularly for weaker diffraction features. At high kinetic energies ($>$ 1000 eV), inelastic scattering is further minimized, and the IMFP increases significantly, enabling sharp forward scattering peaks and well-defined diffraction bands. In this regime, the broadening effect is minimized, and the diffraction pattern approaches the coherent diffraction limit, revealing fine details of the atomic arrangement. So far, only a few studies \cite{fedchenko2020emitter,fedchenko2019high,winkelmann2008high,matsui2012photoelectron,tricot2022energy} address the significant changes in well-defined and faint Kikuchi patterns with kinetic-energy variation, compared to interpreting intensity peaks via azimuthal/polar scans. Even fewer studies have examined circular dichroism asymmetries.\cite{vo2024layered}. 

Via the one-step PES model \cite{vo2024layered}, which can handle the common limitations of other PED codes, this article offers a more complete picture concerning how rich Kikuchi patterns behave at various core levels and are governed by the energy-driven transition. More specifically, we focus on simulating and analyzing fine structures of Si(100) and Ge(100) in a lower range of kinetic energy (106 eV - 4174 eV), as an extension of our earlier effort \cite{vo2024layered}. This energy range corresponds to IMFP values of 0.5 nm to 2.2 nm for Ge and 3.08 nm to 7.61 nm for Si \cite{tanuma2011calculations}, exhibiting both bulk and more 'surface-sensitivity' characteristics. Our simulations successfully replicate experimentally observed bulk-related diffraction features. Previous studies \cite{daimon1993strong,kaduwela1995circular,matsushita2010photoelectron,katayama1999kikuchi,matsushita2005three,tricot2022energy,daimon1993strong} on photoelectron diffraction and circular dichroism in angular distributions (CDAD) for Si have focused only on the 2s and 2p core levels. For the first time, in this work we have observed and simulated the faint Kikuchi patterns of Si 1s, exhibiting CDAD asymmetry of up to 31\%. Furthermore, we provide a detailed investigation of the evolution of well-defined diffraction networks across kinetic energy transitions, in contrast to the more conventional approach of interpreting intensity peaks through azimuthal and polar scans. Under right- and left-circularly polarized light (RCP and LCP, respectively), we offer new insights into their behaviors. The broadening of patterns is firstly discussed for intricate faint features by our model in terms of inelastic scattering treatment.

\section{METHODS}
\subsection{Theoretical model}
The presented \textit{ab-initio} calculations are based on fully relativistic density functional theory, implemented in the multiple-scattering Korringa-Kohn-Rostoker (KKR) Green function-based program (SPRKKR) \cite{ebert2011calculating,ebert2022munich}. The Dirac equation is used to incorporate relativistic effects, such as spin-orbit coupling (SOC). Ground-state calculations for face-centered cubic (FCC) Si (100) and Ge (100) were performed using the experimentally observed lattice constants of 5.431\r{A} and 5.657\AA, respectively \cite{wyckoff1963crystal}. The exchange-correlation potential was approximated using the local density approximation. The potential is treated within atomic sphere approximation (ASA). The self-consistent potential obtained from these calculations was subsequently used for photoemission simulations. The theoretical treatment of core-level photoemission follows the one-step photoemission model, which provides a fully relativistic framework \cite{braun2018correlation}. This approach accurately accounts for all key aspects of ARPES, including the experimental geometry and, crucially, the transition matrix elements that govern dipole selection rules. The final-state wave function is represented by a time-reversed LEED state from Pendry’s
model \cite{pendry1974low,pendry1976theory,hopkinson1980calculation}, allowing the accurate description of these states in a wide range of photon energies (from 6 eV up to several keV) \cite{krempasky2024altermagnetic,strocov2023high,rienks2019large,chaiyachad2024emergence,gray2012bulk,krempasky2016entanglement,krempasky2018operando,beaulieu2020revealing}.

In this framework, the system is modeled as a series of atomic layers, with each layer’s scattering properties described using a partial-wave basis set. Green’s functions are formulated in terms of spherical harmonics and radial functions, enabling the separation of the problem into angular and radial components, where the angular momentum cutoff $l_{max}$ plays a crucial role. The layers are then coupled through a plane-wave basis to construct the solid, with interlayer scattering treated by expanding Green’s functions in an infinite set of reciprocal lattice vectors $\vec{G}_{hkl}$. More details about the implementation can be found in Ref. \cite{vo2024layered}. Practical calculations rely on finite truncations of both basis sets. Determining optimal values for these truncations is nontrivial, as they depend on experimental geometry, photon energy, final-state energy, and atomic composition. Consequently, systematic trial tests must be carefully taken into account. $l_{max}$ = 4 seems to be a fair value for simulations and experiments to agree \cite{vo2024layered}. The inclusion of more $\vec{G}_{hkl}$ vectors results in increasingly complex patterns. This study covers a range of kinetic energy $E_{Final}$ values and various core levels, making it impossible to define a single global number of $\vec{G}_{hkl}$ that ensures optimal agreement. To balance accuracy and computational efficiency, $\vec{G}_{hkl}$ is truncated to a minimum of 45, considering constraints on computational time and memory. The relevant number of
$\vec{G}_{hkl}$ vectors is specified as needed throughout the discussion.

The inelastic scattering corrections to the elastic photocurrent (see Ref. \citet{borstel1985theoretical} or Ref. \cite{braun2018correlation}) is governed by a parametrized and complex inner potential $V_{0}(E) = V_{0r}(E) + iV_{0i}(E)$ \cite{pendry1974low}. This generalized inner potential also accounts for the actual (real) inner potential, which serves as a reference energy inside the solid with respect to the vacuum level \cite{hilgers1995necessity}.  The finite imaginary part $V_{0i}(E_{2})$ of the effective potential for the final state effectively simulates the corresponding inelastic mean free path. As a consequence, the wave field of the excited photoelectron state within the solid can be neglected beyond a certain distance from the surface \cite{pendry1976theory}.

\subsection{Experiment}
The experimental results were measured using the full-field imaging photoemission technique called momentum microscopy (MM). Further details of the experimental techniques are given in Refs.~\cite{tkach2023circular,Medjanik2017}.
Hard X-ray photoemission experiments were conducted on beamline P22 of the PETRA III storage ring at DESY in Hamburg, Germany~\cite{Schlueter2019,Medjanik2019}. 

The footprint of the photon source is approximately 
$50\times 50$~$\mu$m$^2$. A Si (331) crystal monochromator yielded an energy resolution of 150~meV. The single-crystal samples were cut from a wafer and inserted into the sample stage without further cleaning processes.
Circular dichroism experiments in the soft X-ray range were performed
at the soft x-ray ARPES endstation of Beamline I09 at Diamond Light Source, 
UK~\cite{Schmitt2024}.

In this case, the footprint of the photon source is approximately 
$50\times 100$~$\mu$m$^2$. The energy resolution of the grating monochromator was set to 100~meV. The single-crystal samples, also cut from a waver, were cleaned by a HF process and subsequently rinsed by methanol prior to inserting into the sample stage.

\section{RESULTS}
\subsection{Inelastic scattering effects} \label{Inelastic}
\begin{figure}[H]
\centering
\includegraphics[scale=0.4]{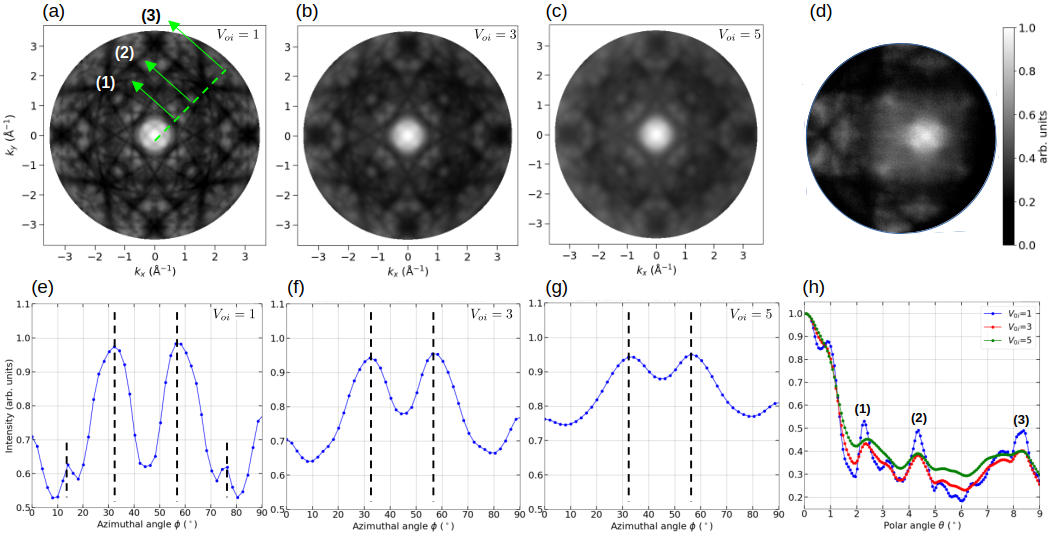}
\caption{(a-c) Calculated and (d) measured total-intensity patterns of Ge(100) 2p\textsubscript{3/2} at $E_{final}$=2017 eV ($h\nu$=3280 eV) as a function of final-state imaginary part $V_{0i}$. (e-g) Azimuthal-angle intensity distributions of photoelectron emission in (a-c) at $\theta = 2.68^{\circ}$, corresponding to the line cut (depicted by a green dash line in (a)). Black dash lines are used to identify the intensity peaks. (f) Polar scan of photoelectron emission in (a-c) at $\phi = 44.25^{\circ}$. The peak (1), (2) and (3) correspond to the ones in (a-c). Calculations are performed with 169 $\vec{G}_{hkl}$. } 
\label{fig:FigA1.png}
\end{figure}

Fig.~\ref{fig:FigA1.png} (a-c) displays the computed diffractograms of Ge 2p\textsubscript{3/2} at the kinetic energy (of electron inside the material) $E_{Final}$=2017 eV ($h\nu$=3280 eV), plotted as a function of the wavevectors $k_{x}$ and $k_{y}$, with respect to the final-state imaginary part $V_{0i}$ (for simplicity $V_{0i} \equiv V_{0i}(E_{2})$. The evolution of simulated photoelectron diffraction patterns provides insight into the role of inelastic scattering. At $V_{0i}$=1 eV (a), the diffractogram exhibits sharp, well-defined Kikuchi bands and intricate diffraction features. Several noticeable similarities to the measured image in Fig.~\ref{fig:FigA1.png} (d) can be identified, such as a bright cross at the center surrounded by a high-intensity circular pattern, bright spots along diagonal directions, and distinct horizontal and vertical Kikuchi lines. As $V_{0i}$ increases, these features progressively blur due to enhanced inelastic scattering. At $V_{0i}$ = 5 eV, the pattern becomes significantly smeared, resembling a diffuse intensity distribution with diminished diffraction contrast. The simulated pattern at higher $V_{0i}$ values closely resembles the experimental diffractogram shown in (d). The loss of fine structure and overall broadening of diffraction features suggest that inelastic scattering is substantial in the measurement, possibly due to finite instrumental resolution, or enhanced electron attenuation. While the central high-intensity region is well reproduced, the visibility of Kikuchi bands and other fine details is reduced. 

The effect of $V_{0i}$ is further investigated in related azimuthal angle scans as illustrated in Fig.~\ref{fig:FigA1.png} (e-g). The polar angle $\theta$ is $13.41^{\circ}$ and the azimuthal angle range is $0^{\circ} \leq \theta \leq 360^{\circ}$ with 180 points. The azimuthal intensity profiles as a function of $V_{0i}$ reveal notable trends in the evolution of diffraction features. A general reduction in intensity contrast is observed, attributed to enhanced inelastic scattering that suppresses coherent diffraction effects. The main peaks (marked by green dashed lines) persist across different $V_{0i}$ values, indicating their robustness against inelastic effects. Their fixed positions remain unchanged with respect to azimuthal angles, which means that they are not significantly altered by the inelastic scattering mechanism. However, these peaks exhibit reduced intensity and noticeable broadening. In contrast, minor peaks (denoted by red dashed lines) disappear and/or merge with surrounding regions with increasing $V_{0i}$, highlighting their sensitivity to inelastic attenuation. The suppression of these secondary features leads to the formation of broader valleys with elevated intensity, suggesting a redistribution of spectral weight. To re-confirm the influence of inelastic scattering, we carry out polar scans Fig.~\ref{fig:FigA1.png} (f) at $V_{0i}$ = 1, 3 and 5. These plots are derived from the cut along $\phi$ = $44.25^{\circ}$ of the diffractograms shown (a), (b) and (c). A sharp intensity drop is observed between $\theta$ = $0^{\circ}$ to $\theta$ = $2^{\circ}$. Beyond this range, three distinct peaks appear, corresponding to three forward-scattering maxima, labeled (1), (2), and (3). The intensity redistribution and peak broadening are evident, consistent with the trends observed in the azimuthal scans. Overall, while the primary diffraction features remain identifiable, increasing inelastic effects smooth out finer oscillations, leading to a more diffuse angular intensity distribution. These findings emphasize the role of inelastic scattering in modifying the photoelectron diffraction pattern and may serve as a useful reference when comparing theoretical predictions with experimental data.

\subsection{Circular dichroism in angular distribution (CDAD) of Si 1s} \label{CDAD_1s}
\begin{figure}[H]
\centering
\includegraphics[scale=0.4]{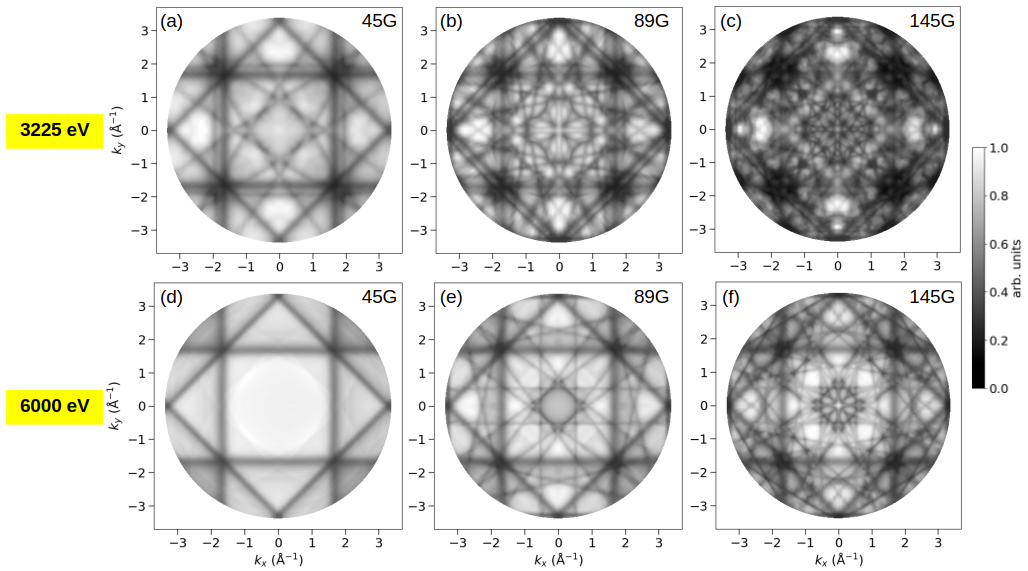}
\caption{Calculated total-intensity patterns of Si(100) 1s at $E_{Final}$ = 1440 eV ($h\nu$ = 3266 eV) (a-c) and $E_{Final}$ = 4174 eV ($h\nu$ = 6000 eV) (d-f) with different numbers of $\vec{G}_{hkl}$ (as mentioned in the panel).} 
\label{fig:FigB1.png}
\end{figure}

Experimentally with computational supports, CDAD existence is verified for adsorbates\cite{westphal1989circular,schonhense1990circular,westphal1991circular,westphal1994orientation,fecher1995solid}, valence-band\cite{schonhense1991circular,schonhense1992circular,ketterl2018origin,nicolai2019bi} and core-level solids\cite{kaduwela1995circular,tkach2023circular,vo2024analyzing}. PED theory offers a comprehensive framework for predicting and analyzing this phenomenon. Before analyzing the CDAD-related attributes, it is crucial to assess the impact of the number of $\vec{G}_{hkl}$ vectors. Fig.~\ref{fig:FigB1.png} presents total-intensity calculations I\textsubscript{TOT} = I\textsubscript{RCP} + I\textsubscript{LCP} for the Si 1s at $E_{Final}$=1440 eV ($h\nu=3266$ eV, first row) and $E_{Final}$=4174 eV ($h\nu=6000$ eV, second row), where I\textsubscript{RCP} and I\textsubscript{RCP} stand for the intensity of RCP and LCP light in turn. To ensure a consistent comparison, all calculations use the same momentum range, resolution, $l_{max}$ value and number of $\vec{G}_{hkl}$. A key observation is that increasing the number of $\vec{G}_{hkl}$ leads to more intricate diffraction patterns, as each set of lattice planes introduces a new "Umklapp channel" \cite{schonhense2020momentum,fedchenko2022structure} (a diffraction pathway associated with Umklapp processes where photoelectron undergoes scattering involving a reciprocal lattice vector $\vec{G}_{hkl}$) in the diffractogram. The diffraction patterns are dominated by a pronounced system of Kikuchi bands, creating a rich fine structure. Each band corresponds to a projection plane flanked by two lines, with its width determined by the associated reciprocal lattice vector.

As the number of $\vec{G}_{hkl}$-vectors increases from 45 to 89 and 145, the fine structure of the diffraction pattern becomes increasingly more complex, with Kikuchi features appearing on an extremely small $\vec{k}$-scale. Unlike conventional electron diffraction, Kikuchi diffraction in PED originates from emitter atoms within the material, which act as sources of the diffracted electron waves. These waves undergo scattering and are diffracted at the crystal's lattice planes before reaching the detector in vacuum. The Kikuchi diffraction mechanism is distinguished by incoherent yet localized electron sources within a crystal. Due to the element-specific binding energies of photoelectrons, PED uniquely enables the identification of chemically distinct emitter sites within a crystal. When measured with sufficiently high angular resolution, photoelectron intensities exhibit a fine structure in their angular distributions \cite{fedchenko2019high}. Experimentally, these filigree structures have not been detected, likely due to the restricted $k$-resolution. Nevertheless, another contributing factor is the reduction of Bragg angles, which increases the real-space path lengths of electron trajectories (see Fig. 3 in Ref. \cite{tkach2023circular}). For high-Z materials like tungsten, the scattering cross section is large and the Kikuchi bands get more and more suppressed when indices rise. Conversely, the observed textures are constrained by the finite number of $\vec{G}_{hkl}$-vectors involved.

The evolution of the pattern when going from $E_{Final}$ = 1440 eV to $E_{Final}$ = 4174 eV is dramatic, in addition to the common behaviors as follows. Overall, the observed band widths and line positions are in good agreement. The fine structures possess mirror symmetry along both the horizontal and vertical planes, aligned with the [010] and [001] crystallographic directions, respectively. The big diamond shape whose edges cross the intersection of vertical and horizontal Kikuchi lines always remains. On the other hand, a striking difference regarding intensity values is captured as the energy rises. Intensities are not always a straightforward summation of intensities from several reflections. Indeed, scattering results in interference terms that produce more complex shapes, such as bright/dark spots or rings, especially when two or more lines cross. The reciprocity principle and the coupling probability of photoelectrons from localized emitters with outgoing waves can be used to understand the bright-dark distribution \cite{winkelmann2008high}. The texture inside the central zone, a square frame created by the intersection of vertical and horizontal low-intensity around the center, evolves significantly via visual inspection. In particular, the bow-tie cross in Fig.~\ref{fig:FigB1.png}b is replaced in Fig.~\ref{fig:FigB1.png}e. However, all geometrical patterns still follow 4-fold symmetry. Obviously, pattern adjustment is forcibly determined by the Bragg angle \cite{kossel1936systematik,de2012structure,winkelmann2019constraints}. As another illustration, four dark spots located at the corners of this central zone shrink. In other words, high-order Kikuchi lines which are thin and possess weak intensities (marked by blue) seem to be invisible at this identical color scale. These minor lines become more prominent at high photon energies (above $\sim$1 keV) as IMFP increases, meaning that electrons travel longer distances before losing energy. As a result, they undergo more elastic scattering events, enhancing multiple scattering contributions. They appear close to the major Kikuchi bands as parallel lines \cite{vespucci2015digital,winkelmann2007many}.

\begin{figure}[H]
\centering
\includegraphics[scale=0.41]{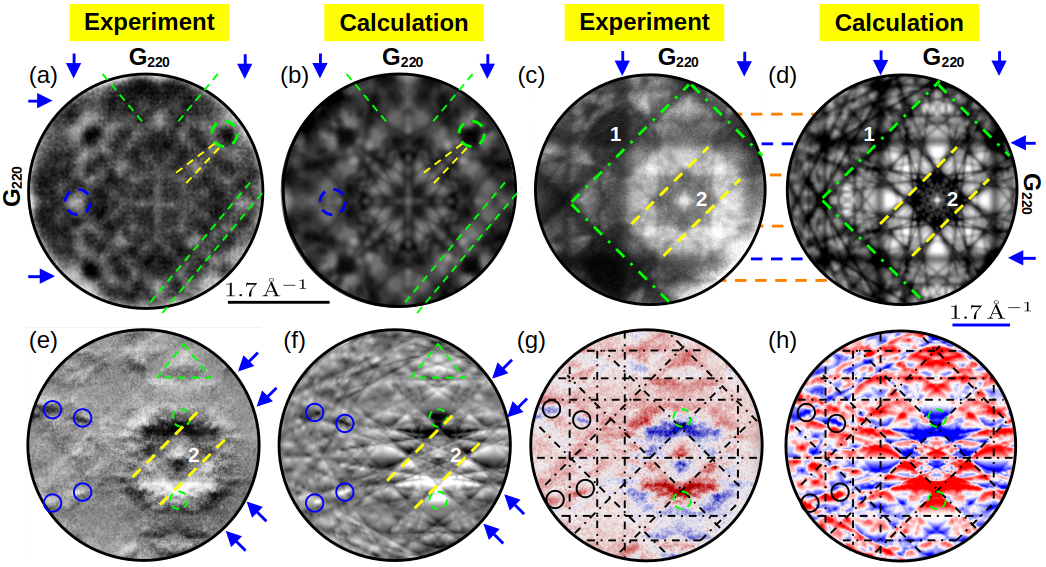}
\caption{Comparison of measured and calculated intensity and CDAD for Si 1s at $E_{Final}$ = 1440 eV ($h\nu$= 3266 eV) and $E_{Final}$ = 4174 eV ($h\nu$= 6000 eV). (a, b) Measured and calculated total intensity pattern I\textsubscript{TOT} at $E_{Final}$ = 1440 eV, respectively; (c, d) Same as (a, b) but at $E_{Final}$ = 4174 eV; (e, f) Measured and calculated intensity difference (CDAD signal) CDAD; (e, i) Measured and calculated CDAD asymmetry A\textsubscript{CDAD}). Computational results are performed with 193 $\vec{G}_{hkl}$. The black and blue scale bar are for the 1s level at $E_{Final}$ = 1440 eV and 3266 eV in turn.} 
\label{fig:FigB2.png}
\end{figure}

Fig.~\ref{fig:FigB2.png} displays a comparison of measured and calculated diffractograms for Si 1s at $E_{Final}$ = 1440 eV ($h\nu$ = 3266 eV) and $E_{Final}$ = 4174 eV ($h\nu$ = 6000 eV). Maximum and minimum intensities are represented by white and black, respectively. The polar angle range of $0^{\circ} \leq \theta \leq 10^{\circ}$ and the whole $360^{\circ}$ azimuthal range ($\phi$) with 180 points for both angles are used to model diffraction. The pattern centers are easily identified without indication. On the left side of the illustration, full arrows (not yet drawn) indicate band edges, while dotted arrows (not yet drawn) indicate band centers. The CDAD signal (also known as intensity difference) is defined as the difference between the intensities of right and left circular-polarized light CDAD = I\textsubscript{RCP} - I\textsubscript{LCP}, and it highlights faint details in the patterns. A so-called CDAD asymmetry quantity A\textsubscript{CDAD} = (I\textsubscript{RCP} - I\textsubscript{LCP})/(I\textsubscript{RCP} + I\textsubscript{LCP}), which reflects the symmetry behavior of CDAD, resulting from normalizing this difference. Because of the autonomy of the spectrometer transmission function and the free-atom differential photoelectric cross-section, this normalized factor is a useful option for experiment-theory comparisons. The A\textsubscript{CDAD} is antisymmetric with respect to the horizontal mirror plane. It is clear that this "up-down antisymmetry" originates the atomic CDAD's symmetry (see Fig. 1 in Ref. \cite{tkach2023circular}) and the Kikuchi diffraction's characteristics (see Fig. 2 in Ref. \cite{tkach2023circular}). To preserve the antisymmetry, the crystal-lattice mirror plane is oriented horizontally. Forward-scattering maxima and circular patterns are derived from a row of atoms, whereas Kikuchi-like bands are produced by Bragg reflections on crystal planes.

The agreement between observed and computed intensity in Fig.~\ref{fig:FigB2.png} (a,b) is quantitatively reasonable. Both patterns exhibit a clear, symmetrical diffraction pattern with a central bright region and repeating features radiating outwards. Four small dark spots (marked as green dash circles) and wedged-shape features (yellow dashes) lying on diagonal lines across the pattern center are captured. Several sets of Kikuchi band edges (indicated by the green dashed lines) are visible in both images, showing good agreement in terms of their position and general orientation. The parallel lines form a big bright diamond shape surrounding a dark smaller one and the pattern center (the bright cross). The angle of the diamond is indicated by a blue dashed circle. These above-mentioned points suggest that the simulation accurately captures the primary diffraction phenomena occurring in the experiment. While the general positions of the Kikuchi lines match, there are differences in the intensity distribution. The experimental image shows broader and more diffuse lines, whereas the calculated pattern exhibits sharper and well-defined lines. This is expected, as the simulation might not perfectly account for all the factors influencing intensity, such as inelastic scattering \cite{herman1991inelastic} and multiple scattering effects \cite{herron2018simulation}. In this study, we do not take into account the contribution from thermal effects \cite{plucinski2008band,papp2011band,braun2013exploring}, leading to an underestimation of the Debye-Waller factor. 
The sharpness of Kikuchi patterns, even in an ideal crystal, is inherently constrained by atomic scattering potentials \cite{winkelmann2010principles,winkelmann2010electron}. The experimental image has a noticeable background noise and some artifacts meanwhile the calculated image is much cleaner and devoid of such noise. The mismatch may derive from the assumption of an idealized bulk crystal, whereas the experimental data are influenced by surface roughness, defects and the limited k-resolution of the electron optics, which resolves approximately 500 points along the image diagonal \cite{fedchenko2020emitter}.

When the photon energy rises to $E_{Final}$= 4174 eV in Fig.~\ref{fig:FigB2.png} (c,d), the overall intensity and contrast is higher than the one from $E_{Final}$ = 1440 eV. A comparison between the experimental and theoretical Kikuchi patterns reveals several key similarities and differences. The overall geometry of the Kikuchi bands (highlighted with blue arrows and green dashed lines) appears in both patterns. The alignment of bright regions (belonging to the band $G_{220}$ between experiment and theory is quite good. The overlapping of the band $G_{010}$ (labeled by yellow dashed lines) and its vertically symmetrized band produces a small dark diamond whose center is a 4-pointed star cross. The pattern marked as (1) is reproduced fairly well in calculations. The CDAD image (Fig.~\ref{fig:FigB2.png}e) appears to have similar patterns to the total intensity image (Fig.~\ref{fig:FigB2.png}c). These include a prominent diamond-shaped structure (labeled '2') at the center, outlined with yellow dashed lines, and a larger, outer diamond that appears incomplete, highlighted with green dashed lines in Fig.~\ref{fig:FigB2.png}c. These similarities are also nicely captured by the one-step photoemission model (Fig.~\ref{fig:FigB2.png}f). Several band edges indicated by yellow dashes and arrows show up in the agreement. Dark-bright features can be mapped accordingly with blue and green indications via the mirror plane. Likewise, these bright/dark shapes emerge from the A\textsubscript{CDAD} patterns as blue/red ones, identified by black and green circles in Fig.~\ref{fig:FigB2.png}(g,h). The consensus between the experimented and simulated A\textsubscript{CDAD} remains highly compelling. The mirror plane and some characteristics (namely Kikuchi bands forming a central diamond) still emerge. To conveniently illustrate the similarities between measuring and computing XPD, the diffractogram is overlaid with a grid of Kukuchi lines (dash dots). Next, we examine the specifics from A\textsubscript{CDAD}. In the upper half of the diffractogram, the red-blue order takes place from the center. The details of 2 rich fine structures are mapped accordingly. Black and green circles show noticeable common features. Others can be obtained by applying the mirror-symmetry operation to the second half. The theoretical A\textsubscript{CDAD} exhibits a rough red-blue texture that closely resembles the experimental counterpart, both in color distribution and relative positioning. The near-perfect agreement confirms the One-step model with forward-scattering approximation as a highly accurate model for Si (Z=14).

Overall, the simulation captures the primary diffraction phenomena occurring in the experiment. Some of the finer details and fainter lines present in the calculated image might be less pronounced or absent in the experimental image due to the aforementioned factors. Another reason originates from the $l_{max}$ effect (above-mentioned in Fig. 4 in Ref. \cite{vo2024layered}. The issue is solvable by increasing $l_{max}$ but for the sake of clearly displaying fine structures in detail, $l_{max}$ =4 is opted. The presence of detailed differences and faint patterns provides strong motivation for improving the current implementation.

\subsection{Bulk Kikuchi photoelectron diffraction} \label{Final_2p}
\begin{figure}[H]
\centering
\includegraphics[scale=0.50]{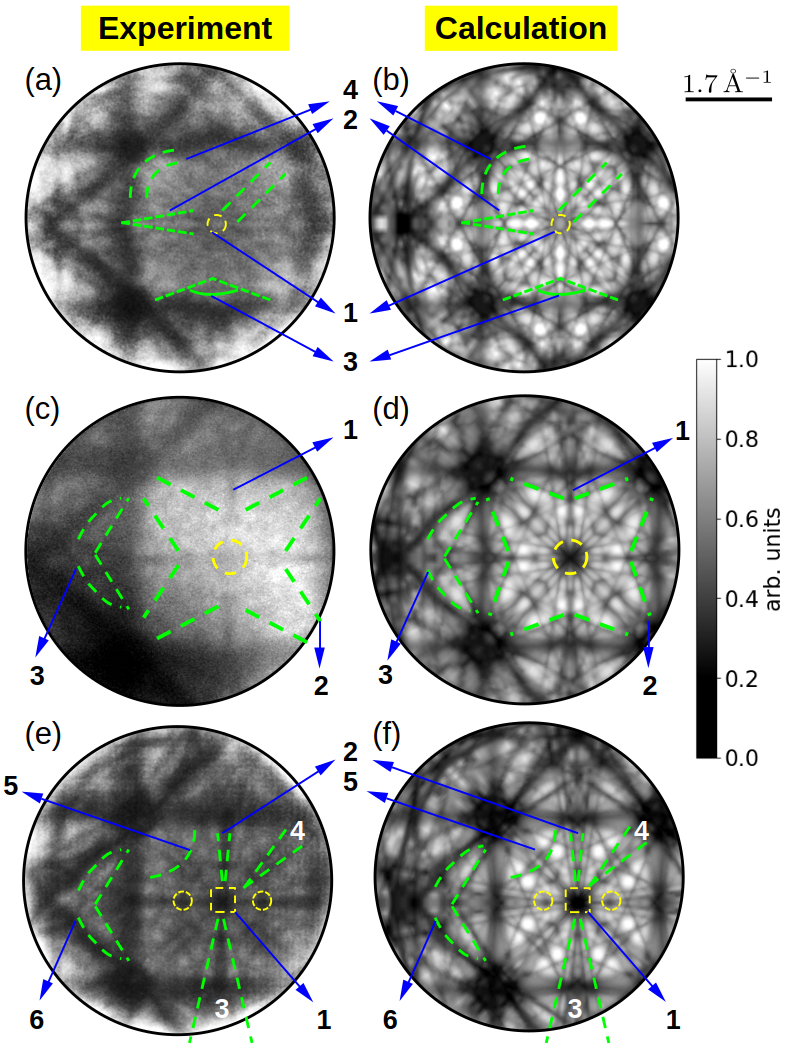}
\caption{Comparison of measured and calculated total intensity pattern I\textsubscript{TOT} for Si 2p\textsubscript{3/2} at $E_{Final}$ = 3180 eV ($h\nu$= 3266 eV; a,b), 3277 eV ($h\nu$ = 3360 eV; c,d) and 3374 eV ($h\nu$ = 3460 eV; e,f). Computational results are performed with 193 $\vec{G}_{hkl}$.} 
\label{fig:FigC1.png}
\end{figure}

In this section, we focus on the 2p core level of Si, building upon a previous work, which explored more bulk-sensitive diffraction patterns (see Figure 1 in Ref. \cite{fedchenko2020emitter}). Fig.~\ref{fig:FigC1.png} shows the results for various $E_{Final}$ = 3180 eV ($h\nu$ = 3266 eV; a,b), 3277 eV ($h\nu$ = 3360 eV; c,d) and 3374 eV ($h\nu$ = 3460 eV; e,f). The patterns display a clear system of Kikuchi bands and lines with a rich fine structure. The experimental and calculated diffractograms match almost perfectly in their overall structure, band widths, and line positions. The only differences are in the relative intensities.  The vertical and horizontal (220)-type Kikuchi bands are particularly striking. The intersection of these bands creates 4 dark spots. A system of dark lines runs throughout the patterns. All patterns possess both horizontal and vertical mirror symmetry.

A detailed examination of (a) and (b) reveals distinctive features indicated by numbered arrows. As eye-catching signatures, we find a dark gray mottle (1) at the center, which is surrounded by four slender triangles (2). Another characteristic is the X-shape (3), whose edges align with diagonal directions. A fourth faint fingerprint is labeled (4), where the tips of these triangles pass through, marked by a larger angle ($>90^{\circ}$). Arcs (5) in the calculations are oriented similarly to those observed in the experiment. This is a clear hint of the 4-fold rotational symmetry. There is a remarkable correspondence between the experimental and theoretical results in all main features.

The second row of Fig.\ref{fig:FigC1.png} (c, d) presents Kikuchi diffractograms at a final-state energy of 3277 eV. Both images exhibit the expected symmetry dictated by the underlying surface crystal structure. A central spot, marked by a yellow dashed circle, is visible in both images and reveals a subtle difference in the intensity gradient between experiment and simulation. Arrow (1) highlights a pronounced intensity minimum following fourfold symmetry, forming a dark cross centered in the image. In the bright central region, arrow (2) points to V-shaped intensity features resembling a “butterfly” pattern. Additionally, a “dolphin-tail” shaped structure, indicated by arrow (3), is clearly present in both diffractograms.

Fig.~\ref{fig:FigC1.png} (e) and (f) show the analogous results for $E_{Final}$ = 3374 eV. Unique features, highlighted by numbered arrows, are apparent upon careful analysis. As eye-catching signatures, we find a dark square mottle (1) at the center, which is surrounded by four slender triangles (2,3). In a specific orientation, two distinct-acute triangles are nested together. However, their tips point toward the center of the pattern rather than outward, as seen in Fig.~\ref{fig:FigC1.png} (a,b). The X-shape is visible in both the measured and calculated diffractograms. Nevertheless, in this case, it is formed by narrow angular features (4) rather than stripes, as depicted in Fig. 1. Arcs (5) emerge in the four diagonal directions in both images. Next to the rim of the Kikuchi band $G_{220}$, we also observe "dolphin-tail" shapes (6), which are clearly reproduced in the calculations as well.

While the positions of key features show good agreement, minor discrepancies in contrast, width, and intensity remain. The calculated patterns display sharper features than the experimental ones. From the experimental side, these differences can be attributed to several factors: suboptimal sample quality, poor cleaving resulting in a "bad spot," slight misalignments (e.g., sample distance or beam footprint), and limited statistics requiring convolution. In contrast, raw simulations are shown to preserve fine details. On the computational side, several approximations may contribute to the deviations. First, using LDA for the exchange-correlation potential may be insufficient \cite{mcgovern1979atomic,wendin1981importance,yuk2024putting}. Second, the ASA treatment imposes shape-related limitations on the potential, which could be mitigated by employing the full-potential SPR-KKR method \cite{minar2005multiple}. Third, our simulations neglect thermal vibrations, conducted at 0 K, while experiments were performed at 30 K. Lastly, real crystals may host imperfections—strain or disorder—not included in our idealized model \cite{westphal2003study,chambers1990elastic}.

\subsection{More "surface-sensitive" photoelectron diffraction} \label{Final_3d}
\begin{figure}[H]
\centering
\includegraphics[scale=0.41]{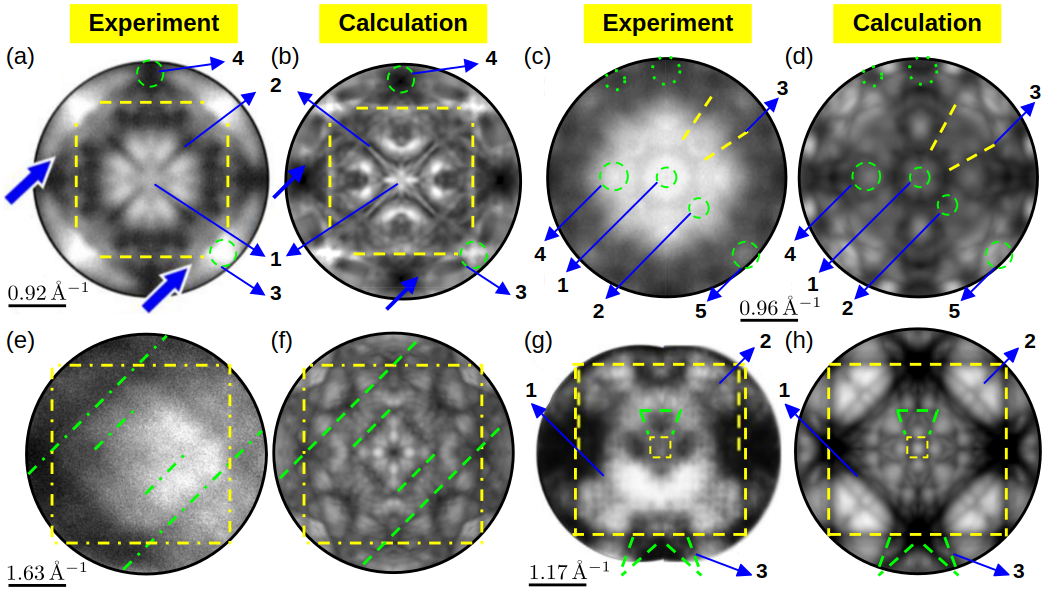}
\caption{Comparison of measured and calculated total intensity patterns I\textsubscript{TOT} for Ge 3d at $E_{Final}$= 106 eV ($h\nu$ = 120 eV; a,b), 586 eV ($h\nu$ = 600 eV; c,d), 787 eV ($h\nu$ = 800 eV; e,f), and 1036 eV (1050 eV; g,h). Computational results at these photon energies are conducted with 45, 69, 97, and 57 $\vec{G}_{hkl}$, respectively.} 
\label{fig:FigD1.png}
\end{figure}

In the following, we span an energy range starting near the minimum of the IMFP curve and extending into the region of bulk sensitivity. Between $E_{Final}$ = 106 and 1036 eV, the IMFP for Ge rises from 0.5 nm to 2.2 nm \cite{tanuma2011calculations,schmitt2024hybrid}, ranging from near the IMFP minimum into the bulk sensitivity regime. Figure ~\ref{fig:FigD1.png} presents the results for different $E_{Final}$ = 106 eV ($h\nu$ = 120 eV; a,b), 586 eV ($h\nu$ = 600 eV; c,d), 787 eV ($h\nu$ = 800 eV; e,f) and 1036 eV ($h\nu$ = 1050 eV; e,f). The patterns exhibit a well-defined system of diffraction lines, enriched with intricate fine structures. The experimental and theoretical diffractograms show good agreement in terms of the overall structure, band widths, and line positions. The patterns follow the four-fold symmetry and possess both horizontal and vertical mirror reflections. However, discrepancies arise in the relative intensities. At E\textsubscript{final} = 106 eV (a,b), the four-point star cross (1) is observed in both images; however, its theoretical size appears smaller. Dark stripes (2) extend along the diagonal directions, forming an X shape. The high intensities at the center of these stripes contribute to discrepancies in the comparison. Additional similarities include bright (3) and dark (4) spots near the edges of the images. These features originate from two squares (indicated by yellow dashes and blue arrows, respectively) that are offset by $45^{\circ}$. Anomalously, we find common characteristics when E\textsubscript{final} = 586 eV (c,d). The central region appears brighter in the experimental data compared to the computational result. This mismatch may stem from image-processing techniques used in the experiments. A gray spot (1) emerges at the center of both diffractograms, surrounded by four smaller dark gray dots (2). Diagonally, four trapezoidal leg-like textures (3) appear, forming an X-shaped pattern that differs from the one observed at E\textsubscript{final} = 586 eV. Other similarities include four high-intensity spots (4) aligned with the vertical and horizontal center lines. The feature labeled (3) in Fig.~\ref{fig:FigD1.png} (a, b) also appears here but with lower intensities (5). Additional resemblances are highlighted by dotted circles.

In Fig.~\ref{fig:FigD1.png} (g,h), the outcome for E\textsubscript{final} = 1036 eV is displayed. "The eye-catching features in both images are the following". The first correspondence is the dark-gray square pattern (marked by yellow dashed) surrounding the diffractogram's center. In calculations, there are four trapezoidal textures (labeled by green dashes) around this square, forming a four-petaled flower motif and corresponding to four dark spots in the measured image. Another similarity is the diamond shape whose edges (1) are quite blurred by measurements. The inner region of this diamond is visibly bright. However, the contrast appears to be higher in (g). Diagonally, a sharp tip (2) pointing away from the center is confirmed. Other fine details include vertical and horizontal lines outlined by the edges (yellow dashes) of another geometric frame, along with 'spot-like' intensity enhancements at its corners. Close inspection brings the appearance of the "double-chevron" structure (3). The majority of the computational data supports experiments whilst misalignment still exists. The latter might come from the crystal lattice imperfection of the sample like defects, dislocations, or impurities. From the theory perspective, more tests with other materials as well as structures are necessary to be conducted in an effort to improve the model.

\section{SUMMARY}
In conclusion, we apply multiple scattering theory and the one-step model of photoemission to analyze Kikuchi diffraction, traditionally studied in electron microscopy. We have systematically investigated the role of inelastic scattering in well-defined Kikuchi patterns for the first time using our model, revealing its effects on intensity redistribution and diffraction broadening. In this work, we have just investigated inelastic scattering as a broadening mechanism. Comprehensive calculations with varying numbers of $\vec{G}_{hkl}$ vectors have identified key factors influencing the fine structure. The prominent features are inherently shaped by diffraction kinematics. We present the first observation and simulation of faint diffraction patterns of Si 1s, revealing CDAD asymmetries of up to 31\%. This study extends previous research, which was limited to the 2s and 2p core levels, providing new insights into the diffraction behavior of deeper core states.  

In addition, we explore a wide range of kinetic energies to understand how diffraction networks evolve under energy-driven transitions. Patterns of the same core level at different energies look different. Our approach successfully reproduces fine and faint structures of bulk and “more surface-sensitivity" diffraction patterns, which is in agreement with the experimental observations. These simulations extend the results of the layer-by-layer method and enable a more quantitative description of photoelectron diffraction (PED) patterns. The computational method presented here can be readily extended to study site-specific effects and complex alloys. By comparing XPD patterns before and after excitation, one can probe the impact of inelastic electron-hole generation and annihilation, which induce both energy loss (gain) and momentum redistribution, fundamentally altering diffraction features \cite{curcio2021ultrafast}. Thus, our study not only underscores the need for a deeper theoretical investigation of intricate faint features in PES but also paves the way for its future extension to ultrafast electronic final-state excitations \cite{greif2016access,ang2020time}.

\section*{Data availability}
Access to the data that support this study is available from the corresponding author upon reasonable request.

\section*{Code availability}
The code utilized for this study is accessible through the corresponding author upon reasonable request.

\section*{Author contributions}
PED reformulations and calculations in the SPRKKR package were performed by T.-P.V. and J.M. T.-P.V. and J.M. conducted the data analysis of the theoretical and experimental PED data with the support from A.P., G.S., H.-J.E., O.T., A.W. and D.S. The experiments were done by O.T., O.F., Y.L., D.V., H.-J.E., and G.S. T.-P.V. wrote the original draft. All authors contributed to the writing, review, and editing of the manuscript.

\section*{Acknowledgement}
This work was supported by the project Quantum materials for applications in sustainable technologies (QM4ST), funded as project \texttt{No. CZ.02.01.01/00/22\_008/0004572} by Programme Johannes Amos Commenius, call Excellent Research (T.-P.V., J.M., A.P.) and the Czech Science Foundation Grant No. GA \v{C}R 23-04746S (T.-P.V.). In addition, T.-P.V., D.S., A.P., and J.M. acknowledge partial funding from Horizon Europe MSCA Doctoral network grant n.101073486, EUSpecLab, funded by the European Union. This work was also supported by the Deutsche Forschungsgemeinschaft, Grant No. TRR288–422213477 (Project B04), and by the Federal Ministry of Education and Research (BMBF, Project 05K22UM2). A.W. was supported by the Polish National Science Centre (NCN), grant no. 2020/37/B/ST5/03669.

\section*{Ethics declarations}
\subsection*{Competing interests}
The authors declare no competing interests.

\bibliography{Reference}
\end{document}